INTERSEZIONI

# Linked open data per la valorizzazione di collezioni culturali: il dataset mythLOD

di Valentina Pasqual e Francesca Tomasi

**Introduzione**
Le tecnologie del web semantico si sono imposte negli ultimi anni come lo strumento per la valorizzazione di collezioni culturali e in particolare come metodo per garantire l'arricchimento dell'esperienza conoscitiva, fondamentale per l'accesso utente. Numerosi sono i database relazionali costruiti da enti e istituzioni culturali che hanno individuato nel web semantico, e nei suoi strumenti teorici, metodologici e tecnici, una soluzione e una opportunità di valorizzazione, nella direzione di una migrazione da modelli relazionali di rappresentazione dei dati a strutture a grafo, capaci di documentare la rete delle relazioni tipizzate. Il grafo è alla base del processo di traduzione del modello concettuale in ontologia, in grado di aumentare il potere espressivo delle collezioni e quindi di consentire l'emergere di nuova conoscenza. Il processo che sovrintende a questo passaggio di strutture di dati però non è neutrale e nemmeno indolore. Una serie di azioni, ormai standardizzate nelle comunità di riferimento, ma pur sempre oggetto di necessaria e attenta definizione, va condotta lungo tutto il percorso di conversione per garantire qualità e valore dei dati finali prodotti[1].

VALENTINA PASQUAL, Alma mater studiorum Università di Bologna, Digital humanities advanced research centre (/DH.arc), Bologna, e-mail valentina.pasqual2@unibo.it.
FRANCESCA TOMASI, Alma mater studiorum Università di Bologna, Digital humanities advanced research centre (/DH.arc), Bologna, e-mail francesca.tomasi@unibo.it.

1 Cfr., in particolare, l'attività del W3C Working Group, *Best practices for publishing linked data*. 2014, <https://www.w3.org/TR/ld-bp/>.

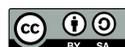

Si intende qui presentare il *workflow*[2] che è stato adottato per la ri-valorizzazione della collezione digitale Mythologiae[3], creata e gestita dal laboratorio di ricerca FrameLAB multimedia & digital storytelling[4] del Dipartimento di beni culturali dell'Università di Bologna. Attraverso l'uso di tecnologie semantiche si è voluto intervenire tanto per garantire la preservazione dei metadati, quanto per formalizzare le esigenze specifiche del dataset, ovvero garantire la massima espressività alla collezione di origine, al contempo arricchendo il potenziale conoscitivo dei dati. Il progetto finale ha portato alla realizzazione della piattaforma mythLOD, che ospita l'intera serie dei contenuti prodotti[5].

Il primo passo necessario alla realizzazione di un qualunque progetto di questo tipo è naturalmente l'analisi dei dati di partenza, perché da tale analisi dipende il risultato del passaggio dal modello relazionale a quello a grafo. Il processo di conversione da formato tabellare a base di conoscenza si è scontrato infatti, nella pratica del progetto, con le classiche problematiche della computabilità delle informazioni che tradizionalmente sono inserite manualmente, quindi, espresse in linguaggio naturale e senza l'ausilio di vocabolari controllati. La procedura adottata in questa fase ha inteso risolvere potenziali ambiguità e indeterminatezze. Attraverso l'uso delle tecnologie del web semantico (RDF e ontologie), e tramite la creazione di entità standardizzate univocamente identificate, si è proceduto a disambiguare e, quando possibile, allineare a fonti esterne (arricchimento di base).

Un secondo passo è quello della definizione dei requisiti. In particolare, nella fase di modellazione dei dati, il progetto si è basato su due aspetti necessari all'ulteriore arricchimento del dataset, in particolare utilizzando strumenti del web semantico nell'ottica della produzione di linked open data (da ora LOD): (i) la formalizzazione dei metadati relativi alle opere museali in Mythologiae, tenendo in considerazione la potenziale soggettività e questionabilità dei dati inseriti; (ii) la formalizzazione delle relazioni (individuate dagli esperti di dominio) tra opere d'arte museali attestate nella collezione d'origine e fonti letterarie correlate. Questi due aspetti identificano il valore aggiunto del *workflow*, che, oltre alla migrazione di formato e di struttura, è la formalizzazione del principio dell'interpretazione come risultato dell'analisi delle fonti di partenza, in direzione di un arricchimento che non sia solo allineamento, ma disvelamento di un nuovo sapere, originariamente basato su una semantica implicita.

La base di conoscenza è quindi rappresentata tanto dai dati di origine della collezione, espressi in LOD, quanto da nuovi dati utili a definire nuovi contesti, necessari a dare consistenza ai dati e ad aumentare il valore conoscitivo del dataset di partenza. Il sistema di collegamenti non è allora limitato ai soli record d'autorità per una riconciliazione, ma si è esteso al livello delle interconnessioni profonde (a semantica esplicita), per la creazione di un *knowledge graph* più espressivo.

Per valorizzare l'intero *workflow*, e volendo in particolare soddisfare l'esperienza dell'utente finale, si è cercato di andare ancora oltre, ragionando anche sulle più adeguate forme di progettazione di interfaccia. Se sul processo di rappresentazione della conoscenza tanto sforzo è stato condotto dalla comunità, molte sono invece le possibilità di riflettere sulle applicazioni per valorizzare l'esperienza utente. Sono state quin-

---

**2** Con il termine *workflow* si intende il flusso di lavoro, organizzato in passaggi sequenziali e iterativi, utilizzato per la migrazione di struttura dati nel progetto mythLOD.

**3** Cfr. <https://patrimonioculturale.unibo.it/mythologiae/>.

**4** Cfr. <https://patrimonioculturale.unibo.it>.

**5** Cfr. <https://dharc-org.github.io/mythlod/static/mima.html>.

di realizzate due visualizzazioni: il catalogo online, arricchito e normalizzato, e la sezione di *storytelling* tematica. In particolare, le due visualizzazioni sono state modellate per soddisfare due requisiti: 1) per essere uno strumento di controllo (verifica della correttezza ed espressività delle entità nella base di conoscenza); 2) per provvedere a un sistema di accesso non convenzionale per l'utente finale (esperti di dominio).

MythLOD è stato sviluppato con alla base l'idea della replicabilità, ovvero in previsione di un riuso del *workflow* in progetti similari, che abbiano il fine della gestione del processo di conversione in LOD. Inoltre, la collezione digitale semantica mythLOD – base di conoscenza, documentazione, visualizzazioni – è istanza della modellazione di un dialogo strutturato tra oggetto museale e fonte letteraria. Per concludere, i materiali conservati nella piattaforma sono rilasciati in formato aperto e sono disponibili anch'essi al reimpiego da parte della comunità.

L'articolo in particolare è organizzato nella seguente modalità: il paragrafo *Stato dell'arte* raccoglie progetti ed esperienze condotte nei vari campi d'interesse relativi al presente studio (in particolare, gestione dei LOD, la creazione di collezioni semantiche e i sistemi di visualizzazione dei dati); la *Metodologia* presenta il *workflow* dell'approccio adottato nella costituzione di mythLOD, dallo studio dei dati all'implementazione delle interfacce; la sezione *Risultati* descrive dataset e relativa documentazione, il catalogo e lo *storytelling*.

**Stato dell'arte**

Le strategie di gestione dei dati del patrimonio culturale in contesto LOD stanno innovando le metodologie tradizionali, stabilendo al contempo nuove modalità di rappresentazione dei dati: modellazione (*modeling*), pulizia o 'bonifica' (*cleaning*), riconciliazione (*alignment*), arricchimento (*enrichment*), collegamento (*linking*) e conseguente pubblicazione (*publishing*) dei dati[6] sono le nuove parole chiave nel campo del dominio GLAM (gallerie, biblioteche, archivi e musei). L'utilizzo delle tecnologie del web semantico[7] come metodo di strutturazione dei dati, consente il riuso di standard di dominio utili a promuovere l'interoperabilità semantica[8].

Sofisticati sistemi di gestione dell'informazione (si possono menzionare, a titolo di esempio, le collezioni semantiche di Europeana[9], o J. Paul Getty Museum[10], British Museum[11], Yale Center for British Art[12], Kunstmera[13], Museo del Prado[14]) ven-gono realizzati dalle istituzioni deputate a conservare, descrivere e valorizzare il patrimonio culturale[15], con lo scopo di rendere i fondi accessibili al pubblico in formato digitale[16]. Il processo di gestione dei dati – dalla sua creazione al rilascio, fino all'uso da parte dell'utente finale – diventa un tema centrale per garantire l'accuratezza e la rappresentatività dell'informazione.

---

[6] Questi i passaggi ritenuti fondamentali in un processo LOD. Per una visione completa ed esaustiva di tale processo, si veda: Seth Van Hooland; Ruben Verborgh, *Linked data for libraries, archives and museums: how to clean, link and publish your metadata*. London: Facet, 2014.

[7] Tim Berners-Lee; James Hendler, *Publishing on the semantic web*, «Nature», 410 (2001), n. 6832, p. 1023-1024, DOI: 10.1038/35074206.

[8] Caroline Sporleder, *Natural Language Processing for cultural heritage domains*, «Language and linguistics compass», 4 (2010), p. 756, DOI: 10.1111/j.1749-818X.2010.00230.x.

[9] Cfr. <https://www.europeana.eu/it>.

[10] Cfr. <https://www.getty.edu/art/collection/>.

[11] Cfr. <https://www.britishmuseum.org/collection>.

[12] Cfr. <https://collections.britishart.yale.edu>.

[13] Cfr. <http://collection.kunstkamera.ru/en/entity/OBJECT>.

[14] Cfr. <https://www.museodelprado.es/en/the-collection/art-works>.

La creazione di una collezione digitale presuppone una serie di passaggi sequenziali, fondamentali e condivisi dalla comunità[17]. Tale processo va affrontato in un'ottica progettuale e trans-disciplinare, riflesso del continuo dialogo tra sviluppatore ed esperto di dominio[18]. Progetti come la collezione dell'Amsterdam Museum[19], l'Archivio della Fototeca Zeri[20], il progetto Agora[21] hanno in comune, tra le attività di gestione dei dati, i passaggi fondamentali di pulizia, riconciliazione, arricchimento, collegamento, pubblicazione, ma dimostrano anche l'importanza dell'adozione di una metodologia articolata, solida, strutturata e ben documentata[22].

In particolare, la pulizia dei dati è un aspetto determinante nel contesto GLAM, considerando la natura strettamente qualitativa del dato. Tecniche di disambiguazione, standardizzazione ed elaborazione del linguaggio naturale (*natural language processing*, NLP) vengono adottate per garantire dati significativi e leggibili, anche attraverso siste-

---

**15** Si pensi all'operato delle tre principali associazioni internazionali, ovvero International Council of Museums (ICOM), International Federation of Library Associations (IFLA) e International Council of Archives (ICA), che trovano nell'Istituto centrale per il catalogo e la documentazione (ICCD), l'Istituto centrale per il catalogo unico (ICCU) e l'Istituto centrale per gli archivi (ICAR) il corrispettivo sul piano nazionale.

**16** Si veda, a titolo di esempio, Martin Doerr, *Ontologies for cultural heritage*. In: *Handbook on ontologies*, Steffen Staab and Rudi Studer (eds.). Berlin; Heidelberg: Springer, 2009, p. 463-486, DOI: 10.1007/978-3-540-92673-3_21.

**17** W3C Working Group, *Best practices for publishing linked data* cit.

**18** Si veda, ad esempio, Francesca Tomasi, *Digital humanities e organizzazione della conoscenza: una pratica di insegnamento nel LODLAM*, «AIB studi», 60 (2020), n. 2, p. 411-425, DOI: 10.2426/aibstudi-12068.

**19** Victor De Boer [*et al.*], *Supporting linked data production for cultural heritage institutes: the Amsterdam Museum case study*. In: *The semantic web: research and applications: 9th extended semantic web conference, ESWC 2012, Heraklion, Crete, Greece, May 27-31, 2012, proceedings*, Elena Simperl [*et al.*] (eds.). Berlin; Heidelberg: Springer, 2012, p. 733-747, DOI: 10.1007/978-3-642-30284-8_56.

**20** Si vedano in particolare: Ciro Mattia Gonano [*et al.*], *Zeri e LODE. Extracting the Zeri photo archive to linked open data: formalizing the conceptual model*. In: *IEEE/ACM Joint conference on digital libraries, London, September 8-12, 2014*, p. 289-298, DOI: 10.1109/JCDL.2014.6970182; Marilena Daquino [*et al.*], *Enhancing semantic expressivity in the cultural heritage domain:exposing the Zeri photo archive as linked open data*, «Journal on computing and cultural heritage», 10 (2017), n. 4, article 21, p. 4, DOI: 10.1145/3051487.

**21** Chie Van Der Akker [*et al.*], *Digital hermeneutics: Agora and the online understanding of cultural heritage*. In: *Proceedings of the 3rd International web science conference, Koblenz, Germany, June 15-17, 2011*. New York: Association for Computing Machinery, 2011, p. 1-7, DOI: 10.1145/2527031.2527039.

**22** Jessica Williams, *Linked Data for libraries, archives and museums: how to clean, link and publish your metadata*, «Collections: a journal for museum and archives professionals», 11 (2015), n. 4, p. 334.

**23** C. Sporleder, *Natural Language Processing for cultural heritage domains* cit., p. 750-753.

mi automatici[23]. Inoltre, l'uso di vocabolari controllati[24], i sistemi di allineamento e riconciliazione delle entità ad altre basi di conoscenza (Wikidata[25], DBpedia[26] ecc.) favoriscono l'interoperabilità e l'arricchimento della base di conoscenza (*knowledge base*), rappresentando al contempo un ulteriore strumento di pulizia.

L'intero processo di gestione dei dati è necessario e fondamentale poi per la (ri)valorizzazione del materiale, ad esempio con intuitivi sistemi grafici (*graphic user interface*), garantendo dunque facile accesso all'utente finale, come lo sono gli esperti di dominio, che non necessariamente sono anche competenti rispetto agli aspetti tecnici e tecnologici[27]. In questa ottica, i ricercatori, che lavorano sulle collezioni, possono potenzialmente trarre un enorme beneficio dalle visualizzazioni, spesso in grado di portare alla risoluzione di complessi problemi di gestione dell'informazione[28].

**Metodologia**

La metodologia è il cuore di ogni processo di trasformazione del sapere. Il progetto mythLOD[29] si basa sulla ri-valorizzazione della collezione digitale Mythologiae, adottando una metodologia che, partendo dall'analisi del progetto d'origine, ha come fine la creazione di una base di conoscenza, capace di rendere contestualmente esplicite le connessioni inespresse, ma presenti nei dati di partenza.

Mythologiae nasce, come anticipato, nel contesto del laboratorio di ricerca FrameLAB e ha l'obiettivo di raccogliere un insieme di fonti iconografiche, classificarle, localizzarle (rispetto agli istituti fisici di conservazione) e infine creare un collegamento fra rappresentazione iconografica e fonti testuali di riferimento. La collezione Mythologiae è il frutto di attività laboratoriali, in cui gli studenti sono stati chiamati a scegliere e interpretare opere d'arte eterogenee, provenienti da diverse realtà di conservazione. Il filo conduttore, che lega fra di loro le opere della collezione, è la scena mitologica che esse raffigurano. Gli annotatori, dunque, non solo hanno compiuto un'analisi ermeneutica di ciascuna opera, assegnando una categoria concettuale (tema ricorrente) concordata sulla base di una tassonomia interna, ma hanno anche associato al tema, dove possibile, le opere letterarie da cui il tema ha tratto origine o quelle opere che lo hanno, in una qualche forma, ripreso.

Un esempio potrà giovare. L'oggetto museale 449 rappresenta la scena di Medea figlicida. La fonte letteraria della scena rappresenta nell'opera viene individuata in "Seneca, Medea, vv. 893-1027". La scena viene poi ripresa in altre opere come "Lenormand Henri, Rene Asie" e "Dolce Lodovico, Medea". Mythologiae raccoglie dunque queste informazioni insieme ai metadati fattuali di ciascuna opera.

---

**24** Antoine Isaac [*et al.*], *Datasets, value vocabularies, and metadata element sets*. 2011, <http://www.w3.org/2005/Incubator/lld/XGR-lld-vocabdataset/>.

**25** Cfr. <https://www.wikidata.org/wiki/Wikidata:Main_Page>.

**26** Cfr. <https://www.dbpedia.org/>.

**27** C. Sporleder, *Natural Language Processing for cultural heritage domains* cit., p. 754-756.

**28** Gary Marchionini; Ryen White, *Find what you need, understand what you find*, «International journal of human computer interaction», 23 (2007), n. 3, p. 205-237, DOI: 10.1080/10447310701702352.

**29** MythLOD contiene i dati convertiti inseriti in Mythologiae fino a giugno 2020.

Al fine della conversione del database Mythologiae in formato LOD, è stato applicato un insieme di passaggi consequenziali e iterativi, ovvero un particolare *workflow* (per cui cfr. Figura 1). Le macrocategorie di tale processo possono essere riassunte in: 1) analisi dei dati di partenza; 2) gestione dei dati - riassumibile in 2.1) modellazione, 2.2) pulizia, collegamento tra entità e produzione del dataset; 3) verifica del dataset e creazione di strumenti di visualizzazione.

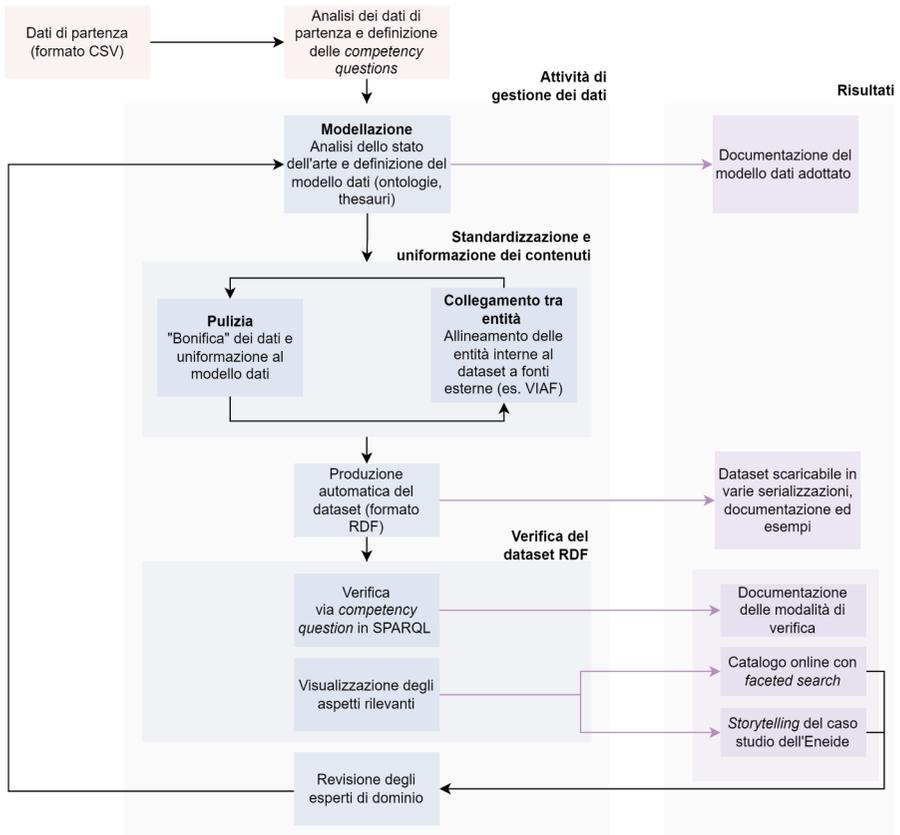

**Figura 1** – Rappresentazione del processo di conversione dei dati (*workflow*) adottato nel progetto mythLOD. In azzurro ne sono rappresentate le attività, in viola i risultati delle attività del progetto

Dall'analisi dei dati di Mythologiae (punto 1 del processo) sono emerse, in particolare, due considerazioni: la presenza di una considerevole quantità di dati qualitativi, ovvero non numerici, ma costituiti da sequenze di stringhe di caratteri; l'assenza dell'impiego di modelli di riferimento, tanto per la struttura (ontologie), quanto per i dati (per esempio vocabolari controllati). Ne è derivato che i dati necessitassero di formalizzazione e della messa in atto di operazioni di revisione, le cui possibili soluzioni sono il risultato di un'analisi approfondita delle singole annotazioni di Mythologiae (cfr. alcuni esempi rilevanti in Figura 2).

| | Metadato | Record in Mythologiae | Definizione del Problema | Definizione della possibile soluzione |
|---|---|---|---|---|
| **Metadati descrittivi dell'opera** | Tipologia dell'opera | a:1:{i:0;s:17:"Pittura vascolare";} | Presenza di "rumore" nel dato, uso di forme eterogee per esprimere lo stesso concetto, mancanza di URI (e/o forme conformi alla creazione dell'URI corrispondente) | Pulizia di "rumore" nei dati, creazione di URI univoci e label standardizzata, esplicitazione di informazioni implicite |
| | Tema mitologico rappresentato dall'opera | medea-figlicida:Medea figlicida | Presenza di "rumore" nel dato, mancanza di URI (e/o forme conformi alla creazione dell'URI corrispondente) | |
| | Identificativo dell'opera | Non presente | Mancanza del dato | |
| **Persone** | Autore dell'interpretazione | Gamba Hubert | Uso di forme eterogee per esprimere lo stesso agente, mancanza di standardizzazione nell'inserimento del nome dell'agente (es. Cognome, Nome) | Creazione di URI univoco per identificare la risorsa e label associata (dove possibile in forma controllata) |
| | Autore dell'opera | Francesco Allegrini | | |
| **Tempo** | Secolo dell'opera | XVII secolo | Formato data non computabile | Traduzione della date in intervalli di tempo e standardizzazione delle stesse in formato machine-readable (ISO) |
| | Anno dell'opera | 1624-1663 | No formato standardizzato | |
| | Data dell'interpretazione | 03/05/2019 07:57 | No formato standardizzato | |
| **Luoghi** | Luogo di conservazione dell'opera | Metropolitan Museum of Art, New York | Agglomerato di informazioni in un solo record (luogo di conservazione + città), uso di forme eterogee per esprimere lo stesso concetto | Creazione di URI univoco per identificare la risorsa con label associata, scissione degli agglomerati di informazioni (istituto conservatore, città, nazione) |
| **Risorse Letterarie** | Fonte Classica associata all'opera | Eneide, IV, 337-396 | Autore implicito, nessuna standardizzazione della citazione, agglomerato di informazioni (opera, libro, verso), uso di forme eterogee per esprimere lo stesso concetto | Scissione di agglomerati di informazioni (autore, opera, verso, libro), esplicitazione di informazioni implicite (e.g. Autore), riuso di forme controllate per identificare le opere e i relativi autori, standardizzazione delle citazioni canoniche tramite URN-CTS |
| | Altre fonti letterarie associate all'opera | Dante, Divina Commedia | No standardizzazione (autore e opera), uso di forme eterogee per esprimere lo stesso concetto | |

**Figura 2** – La tabella rappresenta alcuni campi di metadati di Mythologiae rispetto all'oggetto museale 284. Per ciascun esempio sono presentate le annotazioni in Mythologiae, l'analisi delle problematicità riscontrate e il possibile metodo di risoluzione delle criticità

Con gestione dei dati (2) si intende invece l'intero processo di conversione della collezione Mythologiae da formato tabellare a LOD. Tale processo si è concretizzato in alcune sotto-attività consequenziali e iterative (cfr. ancora Figura 1): modellazione dei dati, ovvero la concettualizzazione del dominio tramite il riuso (e se necessario la definizione) di modelli semantici, nello specifico ontologie, per rappresentare i contenuti di interesse; pulizia e standardizzazione, in funzione della lettura automatica; riconciliazione, o anche controllo d'autorità, mirato a supportare ulteriormente la standardizzazione dei dati (forme controllate dei nomi) e inoltre di favorire l'interoperabilità con altri LOD disponibili sul web; produzione del dataset, ovvero applicazione del modello formale definito in fase di modellazione dei dati e riconciliazione; fase di verifica (*testing*) della base di conoscenza, ovvero analisi e correzione del dataset RDF, prima realizzata attraverso l'uso dei requisiti stabiliti in fase di analisi e modellazione, condotta poi attraverso l'implementazione di due visualizzazioni *ad hoc*. A questi passaggi si è aggiunta la produzione costante di documentazione, necessaria a garantire la maggior trasparenza possibile del processo di gestione dei dati e per favorire il riuso della base di conoscenza.

Ed è in particolare ad alcuni concetti che si vogliono dedicare i prossimi paragrafi; concetti che stabiliscono l'essenza della metodologia utilizzata e la sua dimensione al contempo di riuso delle buone pratiche ma anche di innovazione.

*Modellazione*

La prima attività del processo di gestione dei dati è stata dunque la fase di modellazione (2.1)[30]. Oltre alla definizione dei metadati descrittivi comunemente espressi in ambito museale (ad esempio titolo, autore, istituto conservatore, ecc.), due aspetti sono stati rilevati come fondamentali nell'ottica di ri-valorizzazione delle specificità del dataset di partenza: (i) La formalizzazione esplicita delle interpretazioni degli esperti sugli oggetti museali ("L'opera X rappresenta il tema Y"), mantenendo le informazioni contestuali (l'analisi è un atto interpretativo operato dalla persona P, ed è di tipo iconografico, con criterio ermeneutico); (ii) la formalizzazione delle relazioni assegnate dagli esperti tra opere museali e fonti letterarie associate ("L'opera X rappresenta il tema Y, presente anche nell'opera letteraria W"), anch'esse con le relative informazioni contestuali. L'attività di modellazione dei dati è stata condotta assieme alla definizione dei requisiti per la rappresentazione insieme agli esperti del dominio (FrameLAB). Dopo la revisione dei modelli esistenti, nell'ottica di modellazione di (i) e (ii), si è scelto di applicare il modello di dati *Digital Hermeneutics*[31]. Il modello permette di rappresentare la base di conoscenza su quattro livelli (dati fattuali o *factual data*, asserzioni o *assertion*, informazioni contestuali all'asserzione o *provenance* e alla pubblicazione della stessa o *publication information*), tramite il riuso delle *nanopublication*[32]. Se i primi due livelli del modello sono dipendenti dal dominio di interesse, il secondo e terzo risultano trasversalmente applicabili[33]. Per questo motivo, si è scelto di riusare FRBRoo come ontologia di riferimento per i primi due livelli (dati fattuali e asserzioni) applicati a mythLOD, con l'aggiunta di altri

---

**30** Per maggiori informazioni e dettagli sul modello dei dati di mythLOD cfr. <https://dharc-org.github.io/mythlod/static/datamodel.html>.

**31** Sulla nozione di Digital hermeneutics si veda Marilena Daquino; Valentina Pasqual; Francesca Tomasi, *Knowledge representation of digital hermeneutics of archival and literary sources*, «JLIS.it» 11 (2020), n. 3, p. 59-76, DOI: 10.4403/jlis.it-12642.

**32** Paul Groth; Andrew Gibson; Jan Velterop, *The anatomy of a nanopublication*, «Information services & use», 30 (2010), p. 51-56, DOI: 10.3233/ISU-2010-0613.

**33** M. Daquino; V. Pasqual; F. Tomasi, *Knowledge representation of digital hermeneutics of archival and literary sources* cit., p.73.

campi di metadati descrittivi con Dublin Core Terms[34] e Schema data model[35]. In particolare, considerando (ii), FRBRoo si è presentato come buon candidato per la rappresentazione di un dominio interdisciplinare come quello di mythLOD, il cui interesse risulta trasversale tra contesto museale (le opere d'arte visuali interpretate dagli esperti, al centro della collezione) e bibliografico (la relazione individuata dagli esperti tra gli oggetti museali e fonti letterarie). Nello specifico, le opere d'arte sono state rispettivamente modellate sui due livelli `efrbroo:F2_Expression` e `F4_Manifestation_Singleton` e le risorse letterarie sono state modellate come entità della classe `efrbroo:F1_Work`. Inoltre, tra le risorse letterarie presenti nei metadati ("Riscritture letterarie", "Fonti classiche", "Fonti medievali o moderne", "Riscritture cinematografiche"), le istanze di fonti classiche sono state oggetto di studio ulteriore vista la struttura formale della registrazione delle citazioni canoniche. Le istanze del concetto di `efrbroo:F1_Work` sono state quindi modellate a una granularità profonda tramite il riutilizzo dell'ontologia specializzata HuCit[36], fino ad arrivare al livello della definizione del concetto di citazione canonica (in particolare `hucit:CanonicalCitation`) e riga del testo attraverso l'uso di identificativi URN-CTS[37].

*Pulizia, controllo d'autorità e produzione del dataset*
Il caso di Mythologiae è un classico esempio della problematica già sottolineata da Sporleder[38] rispetto alla gestione di dati non standardizzati: ambiguità linguistica, imprecisioni e concetti sottintesi, comprensibili alla mente umana, ma non alla lettura automatica. Le attività di pulizia dei dati e il controllo d'autorità (2.2) sono state pensate per risolvere tali problematiche. È stato dunque adottato un processo di disambiguazione automatica, semiautomatica e, dove necessario, manuale delle annotazioni di Mythologiae; il processo di standardizzazione di dati ambigui richiede infatti la conoscenza dell'esperto, la quale non può essere completamente automatizzata. Specificamente, il processo di pulizia dei dati si è incentrato sui seguenti aspetti: normalizzazione della punteggiatura, scissione di agglomerati di dati originariamente in un'unica cella, disambiguazione dei contenuti, creazione effettiva di URI non ambigui e univoci, regolarizzazione delle etichette testuali (o possiamo anche dire 'etichetta', cfr. ancora Figura 2). Ad esempio, in Mythologiae la riscrittura letteraria de *I Canti* di Giacomo Leopardi, non seguendo alcuno standard citazionale, è stata registrata in modi diversi (ad esempio "Giacomo Leopardi, Canti", "G. Leopardi, Canti", "G.L. Canti"). È stato necessario dunque ricondurre queste variazioni a un'unica forma controllata. Oppure, ad esempio, annotazioni come "Eneide, IV, 337-396" hanno richiesto l'esplicitazione dell'autore e la divisione logica dei contenuti, allo scopo di creare le entità modellate su HuCit attraverso l'uso di URN-CTS (opera, autore, libro, versi).

---

**34** Cfr. <https://www.dublincore.org/specifications/dublin-core/dcmi-terms/>.

**35** Cfr. <https://schema.org/docs/datamodel.html>.

**36** Si vedano in particolare: Matteo Romanello; Michele Pasin, *Citations and annotations in classics: old problems and new perspectives*. In: *Proceedings of the 1st International workshop on collaborative annotations in shared environment: metadata, vocabularies and techniques in the digital humanities, DH-CASE '13*. New York: Association for Computing Machinery, 2013, p. 1-8, DOI: 10.1145/2517978.2517981 e il collegamento diretto alla risorsa: <https://github.com/mromanello/hucit>.

**37** Cfr. <https://www.homermultitext.org/hmt-doc/cite/cts-urn-overview.html>.

**38** C. Sporleder, *Natural language processing for cultural heritage domains* cit., p. 755-756, p. 758-760.

Il controllo d'autorità è stato utilizzato iterativamente e contemporaneamente alla fase di pulizia dei dati, come ulteriore strumento di disambiguazione dei contenuti. In particolare, OpenRefine[39] è stato utilizzato come strumento di disambiguazione delle annotazioni di partenza tramite la riconciliazione delle stesse a Wikidata. OpenRefine è stato utilizzato in modo semiautomatico per allineare ad altre fonti esterne, metadati come luoghi, periodi, fonti letterarie, autori. Wikidata ha fornito dunque altri dati rilevanti, ad esempio le coordinate geospaziali degli istituti di conservazione degli oggetti museali. Quando possibile, le entità del dataset sono state allineate con VIAF[40] (per autorità e fonti letterarie). Ad esempio, riprendendo l'esempio de *I Canti* di Giacomo Leopardi, l'entità è stata collegata al suo identificativo VIAF (195107635) e alla relativa etichetta controllata ("Leopardi, Giacomo, 1798-1837. | Canti"). Inoltre, ogni citazione canonica è stata allineata con le entità della *HuCit Knowledge Base* (HuCit KB)[41], contenente una selezione di opere classiche modellate con l'ontologia HuCit[42] e il relativo link a Perseus[43]. La base di conoscenza è scaricabile in varie serializzazioni dalla piattaforma mythLOD[44].

*Verifica e creazione di strumenti di visualizzazione*
L'ultimo passaggio del processo di gestione dei dati ha riguardato la validazione della base di conoscenza. Tale processo (3) è stato attuato attraverso due tecniche: (i) definizione dei requisiti per la rappresentazione (*competency questions*) della base di conoscenza, ovvero alcuni quesiti che è necessario porsi per ottimizzare la fase di gestione; (ii) implementazione di due visualizzazioni dei dati. I requisiti definiti con gli esperti di dominio sono stati tradotti in linguaggio SPARQL per interrogare la base di conoscenza prodotta al fine di trovare e correggere errori e al contempo validare l'espressività del modello e dei dati. Di seguito mostriamo un requisito per la rappresentazione espresso in linguaggio naturale che possa essere d'esempio, affiancandolo allo stesso espresso in linguaggio SPARQL (Figura 3) e mostrando i risultati relativi (Figura 4).

Esempio di requisito per la rappresentazione in linguaggio naturale: 'quali sono le fonti letterarie associate al tema Didone e a che tipo di fonte letteraria corrispondono?'

---

**39** Cfr. <https://openrefine.org>.

**40** Cfr. <http://viaf.org/>.

**41** Cfr. <druid.datalegend.net/mromanello/hucit/>.

**42** Matteo Romanello; Michele Pasin, *Using linked open data to bootstrap a knowledge base of classical texts*. In: *Proceedings of the second workshop on humanities in the semantic web (WHiSe II) co-located with 16th International semantic web conference (ISWC 2017)*, edited by Alessandro Adamou, Enrico Daga, Leif Isaksen. CEUR, 2017, p. 3-14, <https://infoscience.epfl.ch/record/232930>.

**43** Si vedano in particolare: David A. Smith; Jeff Rydberg-Cox; Gregory Crane, *The Perseus Project: a digital library for the humanities*, «Literary and linguistic computing», 15 (2000), n. 1, p. 15-25, DOI: 10.1093/llc/15.1.15 e il collegamento diretto alla risorsa: <http://www.perseus.tufts.edu/hopper/>.

**44** Cfr. <https://dharc-org.github.io/mythlod/dataset.html>.

```
PREFIX myth: <https://purl.org/vpq/mythlod/data/>
PREFIX myth-categ: <https://purl.org/vpq/mythlod/data/categ/>
PREFIX efrbroo: <http://erlangen-crm.org/efrbroo/>
PREFIX ecrm: <http://erlangen-crm.org/current/>

SELECT DISTINCT ?work ?type
WHERE {
    GRAPH ?assertion {
        ?work ecrm:P67_refers_to myth-categ:didone }
    GRAPH myth:factual_data {
        ?work a efrbroo:F1_Work ;
            ecrm:P2_has_type ?type }
}
```

**Figura 3** – CQ tradotta in SPARQL

| work | type |
|---|---|
| myth:work/virgil-aeneis | myth:type/fonteClassica |
| myth:work/ungaretti-giuseppe-vita-d-un-uomo | myth:type/riscritturaLetteraria |
| myth:work/alighieri-dante-divina-commedia | myth:type/fonteMedievaleOModerna |
| myth:work/petrarca-francesco-trionfi | myth:type/fonteMedievaleOModerna |
| myth:work/leopardi-giacomo-canti | myth:type/riscritturaLetteraria |
| myth:work/purcell-henry-dido-and-aeneas | myth:type/riscritturaLetteraria |
| myth:work/marmontel-jean-franacois-didon | myth:type/riscritturaLetteraria |

**Figura 4** – La tabella mostra il risultato della CQ indicata

Se, come si è visto, dal punto di vista di un esperto di tecnologie semantiche la base di conoscenza è facilmente interrogabile da qualsiasi SPARQL endpoint, essa non risulta fruibile invece da utenti non esperti del settore. La consultazione dei dati da parte degli esperti di dominio risulta fondamentale per verificarne l'effettiva correttezza[45] e soprattutto la potenzialità espressiva. Nel caso specifico di mythLOD, si aggiunge la mancanza di strumenti che possano visualizzare la base di conoscenza mantenendo la suddivisione nei quattro livelli (cfr. sezione Modellazione).

La visualizzazione della base di conoscenza è stata considerata come un valido strumento di analisi dei dati dal punto di vista della loro correttezza, ma anche, e soprattutto, per verificarne l'espressività. Dunque, i macro-requisiti per la rappresentazione (cfr. paragrafo *Modellazione*, segnalati con la dicitura (i) e (ii)) sono stati riformulati nei termini della visualizzazione lato utente: (i) visualizzazione dei metadati descrittivi degli oggetti museali e delle interpretazioni in mythLOD, mantenendo le informazioni contestuali; (ii) visualizzazione delle relazioni tra opera d'arte e testo con particolare interesse verso le citazioni canoniche nella forma opera, libro, versi.

Sulla base del requisito rappresentativo (i), si è pensato che un catalogo online con sistema di filtri (navigazione a faccette) fosse la soluzione più appropriata. In altre parole, si è pensato di fornire all'utente la possibilità di prendere visione di tutti gli oggetti museali in mythLOD e allo stesso tempo di creare il proprio percorso personalizzato, variabile a seconda degli interessi specifici durante la consultazione del catalogo. Mentre per soddisfare il requisito rappresentativo (ii), tra le opere letterarie presenti nel dataset si è deciso di prendere in esame l'opera virgiliana dell'*Eneide* e di studiare i dati attraverso specifiche visualizzazioni che insieme costituiscono la sezione di *storytelling*.

Se la scelta di creare un'interfaccia intuitiva è dettata dalla necessità di favorire il dialogo con gli esperti di dominio, la scelta degli specifici strumenti di visualizzazione è determinata dalle necessità espressive dei singoli requisiti per la rappresentazione (ovvero dei quesiti potenzialmente formulabili dagli utenti). La visualizzazione dei dati diventa dunque uno strumento strettamente funzionale alla rappresentazione delle specificità del dato, all'esplicitazione del significato semantico ad esso associato (diremo anche della conoscenza) e alla soluzione delle necessità grafiche lato utente, mirando contestualmente a una visione cumulativa della conoscenza registrata nella *knowledge base*, potenzialmente aperta alla scoperta di informazioni e connessioni latenti da parte degli esperti.

---

[45] C. Sporleder, *Natural language processing for cultural heritage domains* cit., p. 754-755.

**Risultati**

Nella metodologia descritta, i risultati delle operazioni di analisi e gestione dei dati garantiscono la produzione di alcuni traguardi intermedi nello svolgimento del progetto, i quali vanno a costituire la documentazione stessa del progetto. Il *workflow* di mythLOD ha prodotto quattro traguardi intermedi (cfr. ancora Figura 1): la base di conoscenza con relativa documentazione, la documentazione del modello dati utilizzato, il catalogo online e lo *storytelling* sull'*Eneide*. Ciascun elemento è reperibile sulla piattaforma mythLOD[46].

La base di conoscenza, prodotto e obiettivo della metodologia applicata, risponde alle problematiche già segnalate in Figura 2. Di seguito, presentiamo la stessa tabella con le soluzioni adottate per la costituzione della base di conoscenza (Figura 5). La tabella esemplifica alcuni metadati prima e dopo la conversione.

| | Metadato | Record in Mythologiae | Record in mythLOD |
|---|---|---|---|
| **Altri Metadati descrittivi dell'opera** | Tipologia dell'opera | a:1:{i:0;s:17:"Pittura vascolare";} | `<https://purl.org/vpq/mythlod/data/type/pittura-vascolare>` rdfs:label "Pittura vascolare"^^xsd:string . |
| | Tema mitologico rappresentato dall'opera | medea-figlicida:Medea figlicida | `<https://purl.org/vpq/mythlod/data/categ/medea-figlicida>` rdfs:label "Medea figlicida"^^xsd:string |
| | Identificativo dell'opera | Non presente | `<https://purl.org/vpq/mythlod/data/item/284>` |
| **Persone** | Autore dell'interpretazione | Gamba Hubert | `<https://purl.org/vpq/mythlod/data/person/gamba-hubert>` rdfs:label "Gamba, Hubert"^^xsd:String |
| | Autore dell'opera | Francesco Allegrini | `<https://purl.org/vpq/mythlod/data/person/allegrini-francesco>` rdfs:label "Allegrini, Francesco, 1729-"^^xsd:String |
| **Tempo** | Secolo dell'opera | XVII secolo | `<https://purl.org/vpq/mythlod/data/time/xvii-secolo>` rdfs:label "XVII secolo"^^xsd:string ; ecrm:P2_has_type `<https://purl.org/vpq/mythlod/data/type/secolo>` ; crm:P82a_begin_of_the_begin "1600-01-01"^^xsd:date ; crm:P82b_end_of_the_end "1699-12-31"^^xsd:date ; |
| | Anno dell'opera | 1624-1663 | `<https://purl.org/vpq/mythlod/data/time/1624-1663>` a ecrm:E52_Time-Span ; rdfs:label "1624-1663"^^xsd:string ; ecrm:P2_has_type `<https://purl.org/vpq/mythlod/data/type/anno>` ; crm:P82a_begin_of_the_begin "1624-01-01"^^xsd:date ; crm:P82b_end_of_the_end "1663-12-31"^^xsd:date . |
| | Data dell'interpretazione | 03/05/2019 07:57 | "2019-05-03T07:57:00"^^xsd:dateTime |
| **Luoghi** | Luogo di conservazione dell'opera | Metropolitan Museum of Art, New York | `<https://purl.org/vpq/mythlod/data/place/the-metropolitan-museum-of-art>` a ecrm:E53_Place ; rdfs:label "The Metropolitan Museum of Art"^^xsd:string ; ecrm:P2_has_type `<https://purl.org/vpq/mythlod/data/type/collocazione>` ; ecrm:P89_falls_within `<https://purl.org/vpq/mythlod/data/place/new-york>`, `<https://purl.org/vpq/mythlod/data/place/united-states-of-america>` ; wdt:P625 "40.77891,-73.96367"^^xsd:string . |

**46** Cfr. <https://dharc-org.github.io/mythlod/static/mima.html>.

| | Metadato | Record in Mythologiae | Record in mythLOD |
|---|---|---|---|
| **Risorse Letterarie** | Fonte Classica associata all'opera | Eneide, IV, 337-396 | `<https://purl.org/vpq/mythlod/data/cit/90>` a hucit:CanonicalCitation ;<br>rdfs:label "Eneide, IV, 337-396"^^xsd:string ;<br>ecrm:P2_has_type `<https://purl.org/vpq/mythlod/data/type/fonteClassica>` ;<br>hucit:has_content `<https://purl.org/vpq/mythlod/data/str/IV-337-396>` ;<br>rdfs:seeAlso "http://data.perseus.org/citations/urn:cts:latinLit:phi0690.phi003.perseus-eng1:4.337-4.396"^^xsd:anyURI . |
| | Altre fonti letterarie associate all'opera | Dante, Divina Commedia | `<https://purl.org/vpq/mythlod/data/work/alighieri-dante-divina-commedia>` a efrbroo:F1_Work ;<br>rdfs:label "Dante, Alighieri (1265-1321) Divina commedia"^^xsd:string ;<br>ecrm:P2_has_type `<https://purl.org/vpq/mythlod/data/type/fonteMedievaleOModerna>` ; |

**Figura 5** – La tabella rappresenta alcuni campi di metadati di Mythologiae rispetto all'oggetto museale 284. Per ciascun esempio sono presentate le annotazioni in Mythologiae e la rispettiva conversione dei dati in formato LOD

Ciascun record di Mythologiae è stato dunque modellato come esemplificato in Figura 5, aggiungendo gli elementi di raccordo necessari per collegare le entità create. La Figura 6 riporta l'annotazione completa dei metadati relativi all'oggetto museale 284, esemplificando una selezione di dati fattuali (*factual data*), l'asserzione relativa all'oggetto (*assertion*), le informazioni contestuali relative all'asserzione (*provenance*) e le informazioni di pubblicazioni (*publication information*) relative ai tre precedenti grafi.

```
@prefix dct: <http://purl.org/dc/terms/> .
@prefix ecrm: <http://erlangen-crm.org/current/> .
@prefix efrbroo: <http://erlangen-crm.org/efrbroo/> .
@prefix crm: <http://www.cidoc-crm.org/cidoc-crm/> .
@prefix hico: <http://purl.org/emmedi/hico/> .
@prefix hucit: <http://purl.org/net/hucit#> .
@prefix myth:   <https://purl.org/vpq/mythlod/data/> .
@prefix np: <http://www.nanopub.org/nschema#> .
@prefix owl: <http://www.w3.org/2002/07/owl#> .
@prefix prov: <http://www.w3.org/ns/prov#> .
@prefix rdfs: <http://www.w3.org/2000/01/rdf-schema#> .
@prefix schema: <http://schema.org/> .
@prefix xsd: <http://www.w3.org/2001/XMLSchema#> .
@prefix co: <http://purl.org/co/> .
@prefix wdt: <http://www.wikidata.org/prop/direct/> .

myth:factual data {
    myth:item/284 a efrbroo:F4_Manifestation_Singleton ;
        ecrm:P2_has_type   myth:type/disegno ;
        ecrm:P55_has_current_location myth:place/the-metropolitan-museum-of-art ;
        efrbroo:R42_is_representative_manifestation_singleton_for myth:item/284-expression ;
        dct:subject "addio"^^xsd:string, "didone"^^xsd:string, "enea"^^xsd:string, "eneide"^^xsd:string ;
        dct:title "La partenza di Enea annunciata a Didone"^^xsd:string ;
        schema:image ""^^xsd:anyURI ;
        rdfs:seeAlso "https://www.metmuseum.org/art/collection/search/338013"^^xsd:anyURI . […]}
```

```
myth:head284 {
    myth:np-284 a np:Nanopublication ;
        np:hasAssertion myth:assertion284 ;
        np:hasProvenance myth:provenance284 ;
        np:hasPublicationInfo myth:pubInfo284 .
}

myth:assertion284 {
    myth:284-expression ecrm:P67_refers_to myth:categ/enea-abbandona-didone.
    myth:cit/90 ecrm:P67_refers_to myth:categ/enea-abbandona-didone .
    myth:work/leopardi-giacomo-canti ecrm:P67_refers_to myth:categ/enea-abbandona-didone . […] }

myth:provenance284 {
    myth:assertion449 prov:wasGeneratedAtTime "2019-05-03T07:57:00"^^xsd:dateTime ;
        prov:wasGeneratedBy   myth:int-act/284 .
    myth:int-act/284 a prov:InterpretationAct ;
        hico:hasInterpretationCriterion myth:sources-association ;
        hico:hasInterpretationType myth:iconographic-approach ;
        prov:wasAttributedTo   myth:person/morelli-martina .
}

myth:pubInfo284 {
    myth:np-284 prov:wasAttributedTo  myth:person/dharc ;
        prov:wasGeneratedAtTime "2020-08-24T09:00:00"^^xsd:dateTime .
}
```

**Figura 6** – Porzione del dataset in serializzazione TriG in riferimento all'oggetto museale 284 rispetto ai quattro livelli concettuali su cui è stato formalizzato

Il dataset RDF (scaricabile in varie serializzazioni), la relativa documentazione del modello formale utilizzato, un caso studio rappresentativo (item 449) di cui è possibile prendere visione di una versione in linguaggio naturale, la resa grafica della porzione del dataset relativa al caso studio, il testo relativo in serializzazione *TriG* (Figura 6) e i requisiti per la rappresentazione tradotti in *query* SPARQL che ne mostrano l'espressività con i relativi risultati, si trovano nella sezione *Knowledge Organisation* del sito.

Le visualizzazioni sono state costruite sulla base dei risultati di query SPARQL specifiche e i risultati sono stati visualizzati con strumenti e librerie[47] *ad hoc*, scelte appositamente per rappresentare il punto di vista d'interesse sui dati.

Le operazioni di visualizzazione dati sono state invece documentate in due sezioni a sé stanti del sito: rispettivamente *Browse the Collection*[48] (catalogo) e *Aeneid Storytelling*[49] (*storytelling*).

Il catalogo mostra una visione globale delle istanze della base di conoscenza, dove, per ogni oggetto museale, è stata realizzata una scheda con i relativi metadati (Figura 7). Ciascuna scheda riporta i metadati dell'opera descritta, raggruppandoli

---

**47** Per la realizzazione della sezione di Storytelling sono stati utilizzati i seguenti tool e librerie: Knight Lab cfr. <https://timeline.knightlab.com/>, Leaflet.markercluster cfr. <https://github.com/Leaflet/Leaflet.markercluster>, Bokeh <https://bokeh.org>, D3.js cfr. <https://d3js.org>.

**48** Cfr. <https://dharc-org.github.io/mythlod/catalogue/>.

**49** Cfr. <https://dharc-org.github.io/mythlod/storytelling/>.

esplicitamente nei livelli del modello dati (dati fattuali, asserzioni, informazioni contestuali). I metadati delle schede nel catalogo sono ripuliti e standardizzati e, dove possibile, collegati a fonti esterne (ad esempio il rimando alla pagina VIAF dell'autore, o il rimando al relativo testo classico in Perseus per le citazioni canoniche - per cui cfr. Figura 7).

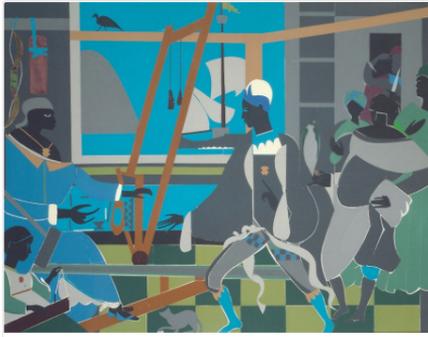

Figura 7 – Scheda del catalogo. La figura mostra l'intera selezione dei metadati visualizzati nell'interfaccia, organizzati secondo tre dei quattro livelli del *Digital Hermeneutics*. L'immagine risulta divisa a metà in senso orizzontale per questioni di spazio

La considerevole quantità di oggetti museali (4260 item in totale) è stata gestita attraverso il già menzionato sistema di filtri (*faceted search*) per una navigazione a faccette, in grado di garantire un ampio spettro di consultazioni della collezione attraverso i campi di metadati considerati particolarmente rilevanti. I livelli del modello dei dati sono stati rappresentati anche in questa sezione al fine di garantire maggior chiarezza (Figura 8). In questo modo, cliccando sul campo dei metadati scelto e successivamente sull'informazione desiderata (ad esempio "Periodo", "Arte Contemporanea"), il sistema attiva il filtraggio delle schede in base alla ricerca effettuata (nell'esempio, 1087 oggetti museali rispettano il canone della ricerca), come mostrato in Figura 8.

**Figura 8** – L'interfaccia del Catalogo online. A sinistra il sistema di navigazione dei metadati con filtro "Periodo" - "Arte Contemporanea" attivati, a destra il risultato della ricerca

Nella sezione di *storytelling* le quattro sezioni (temporalità, localizzazione, concetti chiave e agenti coinvolti) sono visualizzabili organicamente.

La temporalità (*When*) è stata espressa attraverso l'uso di una linea del tempo (Figura 9), che comprende tutti gli oggetti museali, i quali sono stati interpretati e associati all'*Eneide* avendo una scena in comune.

**Figura 9** – Linea del tempo nella sezione di Storytelling in cui è selezionata l'opera d'arte "Didone Costruisce Cartagine" e di cui è possibile visionare alcuni dei relativi metadati

La localizzazione (*Where*) invece è stata tradotta con l'uso di una mappa (Figura 10). Nella mappa sono visualizzati tutti gli oggetti museali che, secondo gli esperti, riprendono una scena tratta dall'*Eneide*.

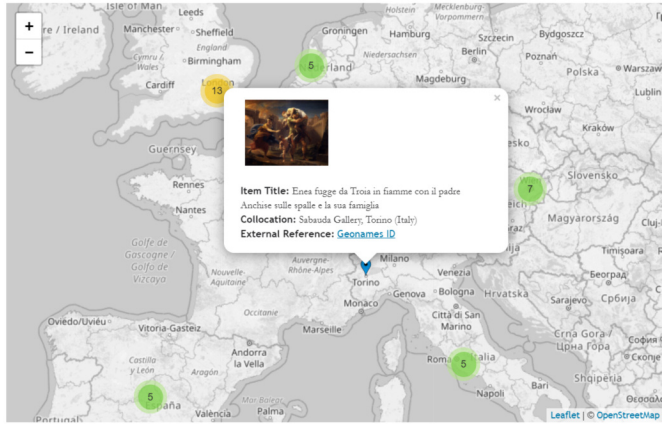

**Figura 10** – La mappa nella sezione di *storytelling* con un oggetto museale selezionato e alcuni dei relativi metadati. In secondo piano si possono vedere i *cluster* di opere museali sparse sulla mappa

I temi mitologici ricorrenti dell'*Eneide* condivisi tra oggetti museali e l'opera virgiliana costituiscono la dimensione del *What* (Figura 11). Questa sezione dello *Storytelling* vuole indagare in modo trasversale la correlazione tra opera museale, tema rappresentato e fonte classica. In particolare, la 'nuvola di parole' (*wordcloud*) mostra i temi rappresentati e le parole chiave degli oggetti museali relazionati all'*Eneide*; due grafici a barre (*barchart*) mostrano gli stessi dati comparando i dieci campi maggiormente istanziati nel dataset.

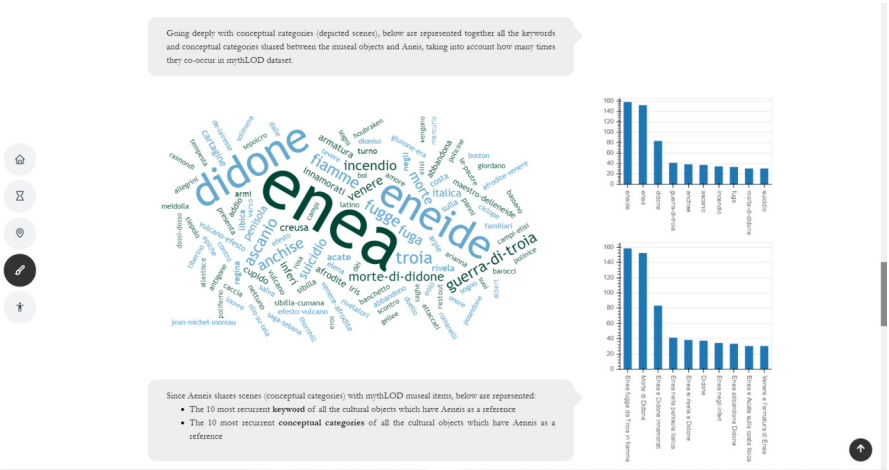

**Figura 11** – Nuvola di parole e grafici a barre nella sezione di *storytelling*

Inoltre, la modellizzazione granulare delle citazioni canoniche ha consentito di costituire una *heatmap* che rappresenta i passi (libri e versi) dell'*Eneide* (Figura 12). Cliccando sulla cella desiderata è possibile visionare quante volte il passo è stato associato a degli oggetti museali e attraverso quale tema.

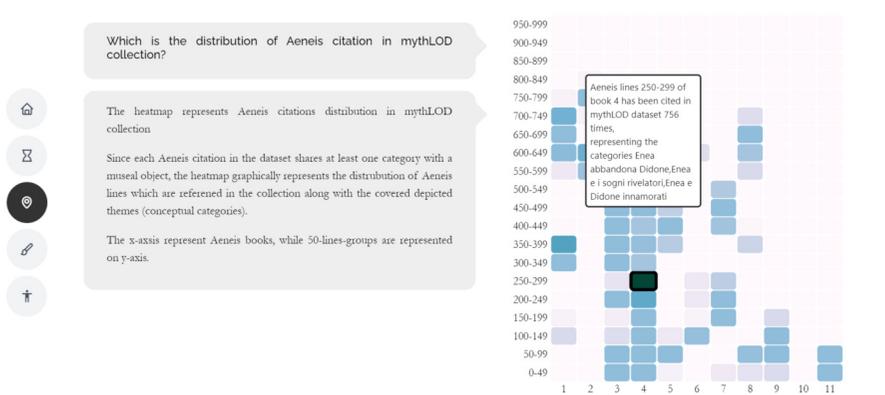

**Figura 12** – *Heatmap* nella sezione di Storytelling con selezione sul IV libro dell'*Eneide*, versi 250-299

Infine, si è voluto mostrare l'insieme delle relazioni che intercorrono nel dataset tra l'*Eneide* e le altre fonti letterarie citate in relazione ai temi ricorrenti che condividono. Per visualizzare questo aspetto si è scelto l'uso del grafo o anche il *network* (Figura 13). Il *network* è stato sfruttato per rappresentare le relazioni tra gli artisti delle opere museali e i temi che essi hanno rappresentato, che sono stati poi riconosciuti dagli esperti del FrameLAB come tratti dall'*Eneide* (*Who*).

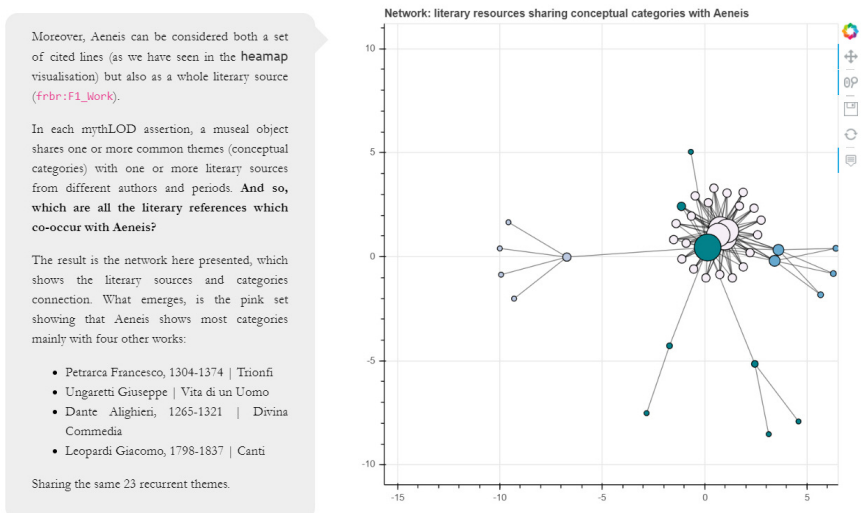

**Figura 13** – Network di relazioni tra fonti letterarie nella sezione di *storytelling* con ingrandimento sui collegamenti tra l'*Eneide* e altre 4 fonti letterarie in relazione a 23 temi ricorrenti

**Conclusioni e sviluppi futuri**

L'articolo ha inteso quindi raccontare l'esperienza della ri-valorizzazione del patrimonio culturale espresso in Mythologiae, esponendo dettagliatamente il *workflow* di conversione da formato tabellare a struttura a grafo. Da questo processo si deduce che la migrazione di formato e di struttura è la formalizzazione del principio dell'interpretazione come risultato dell'analisi delle fonti di partenza, in direzione di un autentico arricchimento, che non voglia essere solo allineamento, ma disvelamento di un sapere nuovo.

Oltre alla tradizionale conversione dei dati, il *workflow* descritto pone inoltre le basi per ragionare sulla rielaborazione dei sistemi di modellazione di interfaccia e di valorizzazione della nuova conoscenza, proponendo un sistema scalabile a largo riuso.

Il modello presentato può essere infatti utilizzato per progetti similari, in particolare nel contesto dell'allineamento tra beni museali e letterari. Il processo di lavoro presentato ha fatto emergere i problemi della classica modellazione dei LOD, definendone i passaggi fondamentali:
- l'analisi dei dati di partenza e definizione dei requisiti per la rappresentazione;
- modellazione nell'ottica non solo della semplice conversione tra formati, ma come strumento di ri-valorizzazione;
- pulizia e collegamento tra entità come strumento per favorire interoperabilità, per la gestione delle forme controllate, ma anche come strumento di arricchimento dell'esperienza utente (ad esempio tramite di link esterni);
- verifica nel contesto della validazione del dataset dal punto di vista sintattico (SPARQL) e semantico-concettuale (tramite la validazione dei contenuti da parte degli esperti di dominio attraverso le visualizzazioni).

In particolare, la piattaforma mythLOD è una effettiva testimonianza dei benefici dell'uso dei LOD nel contesto dei beni culturali, presentandosi come un percorso necessario per connettere elementi e usufruire di queste connessioni in termini di arricchimento dei dati. L'utilizzo delle buone pratiche descritte nel *workflow* è confermato dalla base di conoscenza di mythLOD e in particolare dalle due visualizzazioni. Il catalogo è navigabile attraverso un sistema di filtri basato sulla navigazione a faccette relativa a metadati delle opere d'arte, fonti letterarie e informazioni contestuali degli atti interpretativi. Inoltre, i dati delle schede sono collegati a fonti esterne, fornendo, ad esempio, la possibilità di visualizzare un passaggio specifico di un'opera classica visualizzabile su Perseus.

L'applicazione del *workflow* risulta necessaria anche alla creazione della sezione di *storytelling*, che permette di valorizzare tanto i dati temporali, utilizzati per la creazione della linea del tempo, quanto le coordinate geospaziali dei luoghi, per visualizzare le istanze sulla mappa. Inoltre, la sezione di *storytelling* mostra come la migrazione di formato da tabella a LOD permetta all'utente di spostare il punto di vista dalla collezione (soggetto del catalogo) alle fonti letterarie (l'*Eneide* come soggetto dello *storytelling*).

La visualizzazione, come dimostrato, è allora un esempio di come impiegare procedure mirate a valorizzare il ruolo dell'utente finale nell'accesso alle risorse. Essa rientra fra quelle azioni di ricerca che vogliono vedere nei LOD una nuova modalità non solo di pubblicazione, ma anche di semantizzazione del patrimonio culturale, ovvero di arricchimento dell'esperienza di accesso, nell'ottica di elaborare nuove forme di organizzazione, e quindi di fruizione, della conoscenza.

VALENTINA PASQUAL, Alma mater studiorum Università di Bologna, Digital humanities advanced research centre (/DH.arc), Bologna, e-mail valentina.pasqual2@unibo.it.
FRANCESCA TOMASI, Alma mater studiorum Università di Bologna, Digital humanities advanced research centre (/DH.arc), Bologna, e-mail francesca.tomasi@unibo.it.

**Linked open data per la valorizzazione delle collezioni dei beni culturali: la produzione del *dataset* mythLOD**
La rappresentazione formale dei metadati culturali è sempre stata una sfida, considerando tanto l'eterogeneità degli oggetti culturali quanto l'esigenza di documentare l'atto interpretativo esercitato dagli esperti. Questo articolo presenta una panoramica della ri-valorizzazione della collezione digitale Mythologiae in formato linked open data. La ricerca mira a esplorare i dati di una raccolta di opere d'arte (Mythologiae) promuovendo le potenzialità del web semantico, concentrandosi in particolare sulla rappresentazione formale dell'associazione degli oggetti culturali alle fonti letterarie, così come realizzata dagli esperti, documentando anche le loro interpretazioni.
Il flusso di lavoro è consistito nella definizione del modello di dati, nella pulizia e disambiguazione degli stessi, nella conversione (da struttura tabulare a grafo) e nell'attività di *testing* (in particolare la revisione degli esperti di dominio del *dataset* tramite *competency question* e visualizzazioni dei dati). Il risultato è la piattaforma mythLOD, che presenta il *dataset* insieme alla documentazione dettagliata della ricerca. Inoltre, la piattaforma ospita i due spazi di visualizzazione dei dati (il catalogo online e un esperimento di *data-storytelling* sul caso studio dell'*Eneide*) che arricchiscono la documentazione del progetto come unità di test *user-friendly* per il *dataset* e costituiscono un ulteriore strumento di documentazione del progetto e di esplorazione della collezione.

**Enhancing cultural heritage collections through linked open data: the production of the mythLOD dataset**
The formal representation of cultural metadata has always been a challenge, considering the heterogeneity of cultural objects and especially when dealing with experts' interpretation over them. This paper presents an overview of the mythLOD dataset production as the Mythologiae digital collection revalorisation into linked open data format. The research aims then to explore Mythologiae data leveraging semantic web potentialities, focusing over the formal representation of experts' analysis when associating visual artworks (and their interpretations) to literary sources.
The workflow of the project consisted of data model definition, data cleaning and entity linking, conversion (from tabular data to graph) and testing activity (domain experts review over competency questions and two designed on-purpose data visualizations). The result is the mythLOD platform, which present the dataset and the detailed documentation of the research. Additionally, the platform hosts the two data visualization spaces which have been implemented – the online catalogue and the storytelling experiment over *Aeneid* – as a user-friendly testing unit for the dataset and an additional tool for documenting the project and exploring the collection.